\documentclass[fleqn,usenatbib]{mnras}

\usepackage{newtxtext,newtxmath}

\usepackage[T1]{fontenc}

\DeclareRobustCommand{\VAN}[3]{#2}
\let\VANthebibliography\thebibliography
\def\thebibliography{\DeclareRobustCommand{\VAN}[3]{##3}\VANthebibliography}

\usepackage{graphicx}	
\usepackage{amsmath}	
\usepackage{amssymb}	

\usepackage{subcaption} 
\usepackage{upgreek} 
\usepackage{xspace} 
\usepackage{color, colortbl} 
\usepackage{gensymb} 
\usepackage{soul} 

\newcommand{\cii}{[C\,{\sc ii}]\xspace}
\newcommand{\oiii}{[O\,{\sc iii}]\xspace}
\newcommand{\olong}{[O\,{\sc iii}] 88\,$\upmu$m\xspace}
\newcommand{\oshort}{[O\,{\sc iii}] 52\,$\upmu$m\xspace}
\newcommand{\hii}{H\,{\sc ii}\xspace}

\newcommand{\nii}{[N\,{\sc ii}]\xspace}

\title{A solar metallicity galaxy at \(z>7\)? Possible detection of the \nii 122\,$\upmu$m and \oshort lines}

\author[M. Killi et al.]{
Meghana Killi,$^{1,2}$\thanks{E-mail: meghana.killi@nbi.ku.dk}
Darach Watson,$^{1,2}$
Seiji Fujimoto,$^{1,2}$
Hollis Akins,$^{3}$
Kirsten Knudsen,$^{4}$
Johan Richard,$^{5}$
\newauthor Yuichi Harikane,$^{6,7}$
Dimitra Rigopoulou,$^{8,9}$
Francesca Rizzo,$^{1,2}$
Michele Ginolfi,$^{10}$
Gergö Popping,$^{11}$
\newauthor Vasily Kokorev$^{1,2}$
\\
$^{1}$Cosmic Dawn Center (DAWN)\\
$^{2}$Niels Bohr Institute, University of Copenhagen, Jagtvej 128, 2200 Copenhagen N, Denmark\\
$^{3}$Department of Astronomy, The University of Texas at Austin, 2515 Speedway Blvd Stop C1400, Austin, TX 78712, USA\\
$^{4}$Department of Space, Earth and Environment, Chalmers University of Technology, Onsala Space Observatory, Onsala, SE-43992, Sweden\\
$^{5}$Univ Lyon, Univ Lyon1, ENS de Lyon, CNRS, Centre de Recherche Astrophysique de Lyon UMR5574, 69230 Saint-Genis-Laval, France\\
$^{6}$Institute for Cosmic Ray Research, The University of Tokyo, 5-1-5 Kashiwanoha, Kashiwa, Chiba 277-8582, Japan\\
$^{7}$Department of Physics and Astronomy, University College London, Gower Street, London WC1E 6BT, UK\\
$^{8}$Astrophysics, Department of Physics, University of Oxford, Keble Road, Oxford OX1 3RH, UK\\
$^{9}$School of Sciences, European University Cyprus, Diogenes Street, Engomi, 1516 Nicosia, Cyprus\\
$^{10}$Dipartimento di Fisica e Astronomia, Università di Firenze, Via G. Sansone 1, 50019, Sesto Fiorentino (Firenze), Italy\\
$^{11}$European Southern Observatory, Karl-Schwarzschild-Str. 2, D-85748, Garching, Germany
}

\date{Accepted XXX. Received YYY; in original form ZZZ}

\pubyear{2022}

\begin{document}
\label{firstpage}
\pagerange{\pageref{firstpage}--\pageref{lastpage}}
\maketitle

\begin{abstract}
We present the first detection of the \nii 122\,$\upmu$m and \oshort lines for a reionisation-epoch galaxy. 
Based on these lines and previous \cii 158\,$\upmu$m and \olong measurements, using two different radiative transfer models of the interstellar medium, we estimate an upper limit on electron density of \(\lesssim500\)\,cm\(^{-3}\) and an approximate gas-phase metallicity of \(Z/Z_{\sun} \sim 1.1\pm0.2\) for A1689-zD1, a gravitationally-lensed, dusty galaxy at \(z = 7.133\).
Other measurements or indicators of metallicity so far in galaxy interstellar media at \(z\gtrsim6\) are typically an order of magnitude lower than this.
The unusually high metallicity makes A1689-zD1 inconsistent with the fundamental metallicity relation, although there is likely significant dust obscuration of the stellar mass, which may partly resolve the inconsistency. 
Given a solar metallicity, the dust-to-metals ratio is a factor of several lower than expected, hinting that galaxies beyond \(z\sim7\) may have lower dust formation efficiency. 
Finally, the inferred nitrogen enrichment compared to oxygen, on which the metallicity measurement depends, indicates that star-formation in the system is older than about 250\,Myr, pushing the beginnings of this galaxy to \(z>10\).
\end{abstract}


\section{Introduction}

The frontier of the study of galaxy evolution has now moved to the epoch of reionisation, \(z \gtrsim 7\), where the physical conditions of the interstellar medium (ISM) are beginning to be investigated \citep[e.g.][]{novak_alma_2019,bouwens_reionization_2022}. Measuring these conditions is critical to our understanding of the evolution of galaxies and the growth of structure. The metal enrichment of the gas in galaxies, in particular, can tell us about the extent of processing of the ISM through stars, and therefore, the stage of evolution of the galaxy. However this fundamental ISM property is difficult to determine at high-\(z\).

While at low redshifts, ISM properties are often determined using optical and ultraviolet (UV) emission lines \citep[e.g.][]{kewley_understanding_2019,maiolino_re_2019}, at \(z>7\) those lines shift into the infrared (IR), where \emph{JWST} is just beginning to produce the first results \citep[e.g.][]{schaerer_first_2022, curti_chemical_2022}. However, heavily dust-obscured galaxies \citep[e.g.][]{marrone_galaxy_2018,fudamoto_normal_2021} cannot be studied with \emph{JWST} because UV-optical observations cannot probe dust-obscured gas \citep{chartab_gas_2022}. Hence, at high redshift, we require detections of bright far infrared (FIR) cooling lines and dust emission to estimate ISM properties \citep[e.g.][]{nagao_metallicity_2011,novak_alma_2019}.

So far, FIR lines such as \cii\,158\,$\upmu$m and \olong (hereafter [C158] and [O88] respectively) have been detected in only a handful of \(z>7\) galaxies \citep[e.g.][]{maiolino_assembly_2015,pentericci_tracing_2016,carniani_extended_2017,hashimoto_big_2019,carniani_missing_2020, sommovigo_dust_2021, schouws_alma_2022}. Very few galaxies have been detected in both [O88] and [C158] at \(z>6\) \citep{carniani_extended_2017,hashimoto_big_2019,tamura_detection_2019,bakx_alma_2020,harikane_large_2020, witstok_dual_2022}, and only four of those are at \(z>7\).

Furthermore, while observations of the [O88] and [C158] lines and continuum emission allow the star-formation rate, dust mass and, to some extent, the temperature to be assessed with some reliability, determining the basic ISM parameters, i.e.\ the gas-phase metallicity, density, and ionisation parameter, requires other FIR lines. For instance, \citet{pereira-santaella_far-infrared_2017} and \citet{harikane_large_2020} describe models that use lines such as \oshort and \nii\,122\,$\upmu$m (hereafter [O52] and [N122] respectively) in addition to [O88] and [C158].

However, this poses an observational challenge because while [O88] and [C158] are bright, [N122] is relatively faint and difficult to detect at \(z > 6\). There have been [N122] detections in quasar host galaxies at \(z = 6.003\) \citep{li_ionized_2020}, and \(z=7.54\) \citep{novak_alma_2019}, but non-detections for all other systems attempted at \(z \sim 6\text{--}7\) \citep{harikane_large_2020,sugahara_big_2021}.
Although the [O52] line can be bright, it is also difficult to detect at this redshift as it lies in a wavelength region with low atmospheric transmission. Thus far, there have been no detections reported of [O52] at \(z > 6\).

In this work, we report on the first measurement of the [O52] and [N122] lines for a non-quasar galaxy at \(z > 6\). Together with previous [O88] and [C158] measurements (\citealt{akins_alma_2022,wong_alma_2022}; Knudsen et~al.\ (in prep.)), we now have four FIR line detections for the gravitationally-lensed reionisation-epoch, dusty, normal galaxy A1689-zD1 at \(z = 7.133\), making it the ideal candidate to study ISM conditions in re-ionisation era galaxies. 

A1689-zD1 is lensed by the galaxy cluster Abell\,1689 with a magnification factor of 9.3 \citep{watson_dusty_2015}. It was first discovered as a photometric candidate \(z>7\) galaxy \citep{bradley_discovery_2008}. The Ly$\alpha$ break was spectroscopically confirmed with deep VLT/X-shooter data, and it was shown to be a dusty galaxy with ALMA detections in bands 6 and 7 \citep{watson_dusty_2015,knudsen_merger_2017}. This was the first detection of dust at \(z>7\), though more distant dust emitters have since been identified \citep[e.g.][]{fudamoto_normal_2021, ferrara_alma_2022, schouws_significant_2022,laporte_dust_2017}. A1689-zD1 has now been detected in strong [C158] and [O88] emission \citep{wong_alma_2022}, the detailed 2D and 3D structure of which is studied in \citet{akins_alma_2022} and Knudsen et~al.\ (in prep.). The galaxy has also been detected in four continuum bands allowing an accurate measurement of its dust temperature and mass \citep{bakx_accurate_2021}. The rich multi-wavelength data set makes it one of the best-studied reionisation-epoch galaxies.

In this paper, we report the measured line fluxes, and calculate ratios among the four lines and their underlying continua to characterise the ISM of A1689-zD1. We deal here mainly with the galaxy-integrated properties. A resolved study of A1689-zD1 is presented in Knudsen et~al.\ (in prep.).

We adopt a Flat $\Lambda$CDM cosmology with $\rm H_0=67.74\,km\,s^{-1}\,Mpc^{-1}$, $\rm \Omega_M=0.3075$, and $\rm \Omega_\Lambda=0.6925$ \citep{planck_collaboration_planck_2016}.

\section{Observations and Methods}

For the analysis presented in this work, we use the following the values for the stellar mass ($M_{\rm \ast}$), dust mass ($M_{\rm d}$), and star-formation rate (SFR) for the galaxy: \(\mathnormal{M}_\ast = 1.7^{+0.7}_{-0.5} \times 10^9 \mathrm{M}_\odot\) \citep{watson_dusty_2015}; total SFR \( = 37 \pm 1\, \mathrm{M}_\odot\)\,yr\(^{-1}\) \citep{akins_alma_2022}; \(\mathnormal{M_{\rm d}} = 1.7^{+1.3}_{-0.7} \times 10^7 \mathrm{M}_\odot\) \citep{bakx_accurate_2021}.

\subsection{[N122] and [O52] observations}

Observations were carried out at the Atacama Large Millimeter/submillimeter Array (ALMA) in Chile from November to December 2019 in cycle\,7 (\# 2019.1.01778.S, PI: D. Watson) under a precipitable water vapour (PWV) of 0.3--0.8\,mm, using 42--45 antennae with projected baselines of 15--313\,m. 
Based on a source redshift of \(z=7.1332\pm0.0005\), securely determined with [C158] and [O88] (\citealt{wong_alma_2022}, Knudsen et~al.\ (in prep.)), the available 7.5\,GHz bandwidth with four spectral windows was centred at observed frequencies of 296.9\,GHz (Band~7) and 703.8\,GHz (Band~9) so that the [N122] and [O52] lines fall in one or two spectral windows.    
J1229+0203 and J1337$-$1257 were observed as the flux and bandpass calibrators. Phase calibration was performed by using observations of J1256$-$0547. 
The total on-source times were 200\,minutes and 95\,minutes for the [O52] and [N122] observations respectively.


We reduced the ALMA data with the Common Astronomy Software Applications package (CASA; \citealt{mcmullin_casa_2007})   
in the standard manner with the scripts provided by the ALMA observatory. 
We produced the continuum images and line cubes by running the CLEAN algorithm with the {\sc tclean} task.
For continuum, 
we flagged the calibrated visibility in the expected frequency ranges of the lines.
We executed the {\sc tclean} routines down to the \(1\sigma\) level with a maximum iteration number of 10\,000 in the automask mode with the sub-parameters determined by the recommendations of the ALMA automasking guide\footnote{https://casaguides.nrao.edu/index.php/Automasking\_Guide}. 
For cubes, 
we applied continuum subtraction to the calibrated visibility with the {\sc uvcontsub} task by using the line-free frequency. We fit the continuum along channels at least ±500 km/s away from the expected line centre. We tried the subtraction with fitorders 0, 1, and 2. For [O52], the automasking and cleaning worked best for fitorder 0, and for [N122] the results were similar for all fitorders. We therefore chose to use fitorder of 0 for both [O52] and [N122] continuum subtraction.
We adopted a spectral channel width of 20 km\,s$^{-1}$ and performed the CLEAN algorithm in each channel in the same manner as the continuum map.
In both cases, we used natural weighting to maximise the sensitivity and applied the multi-scale deconvolver with
scales of 0 (i.e. point source), 1, and 3 times the beam size.
We list the synthesised beam size and the standard deviation of the pixel values in the final natural-weighted maps and cubes in Table~\ref{tab:data}. 


\begingroup
\setlength{\tabcolsep}{3.5pt}
\begin{table}
	\caption{Data properties for the four lines and two continuum images used in this work. The [O52] and [N122] data were produced by natural weighting, while the [O88] and [C158] data use Briggs weighting as described in Knudsen et~al.\ (in prep.). Beam size corresponds to the full width at half maximum (FWHM). The \(1\sigma\) sensitivity is evaluated from the standard deviation of the pixel values.}
	\label{tab:data}
    \begin{tabular}{@{}cccc@{}} 
    \hline\hline
		ALMA Band & Target line/continuum & Beam-size & Sensitivity \\
    		      &           & [$''\times''$] & [mJy\,beam$^{-1}$] \\
    \hline
		6 & [C158] & $0.24\times0.22$ & 0.02 \\
		7 & [N122] & $1.19\times0.98$ & 0.02 \\
		8 & [O88] & $0.33\times0.28$ & 0.12 \\
		9 & [O52]  & $0.51\times0.42$ & 0.42 \\[3pt]
		7 & continuum & $1.18\times0.97$ & 0.06 \\
		8 & continuum & $0.46\times0.40$ & 0.05 \\
    \hline
	\end{tabular}
\end{table}

\section{Results}
\label{sec:results}

\subsection{Detection}

\begin{figure*}
\centering
    \includegraphics[width=0.1\textwidth]{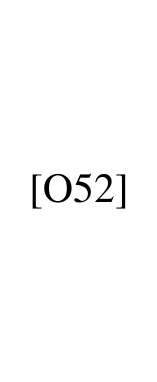}\hfill%
    \includegraphics[height=0.22\textheight, keepaspectratio]{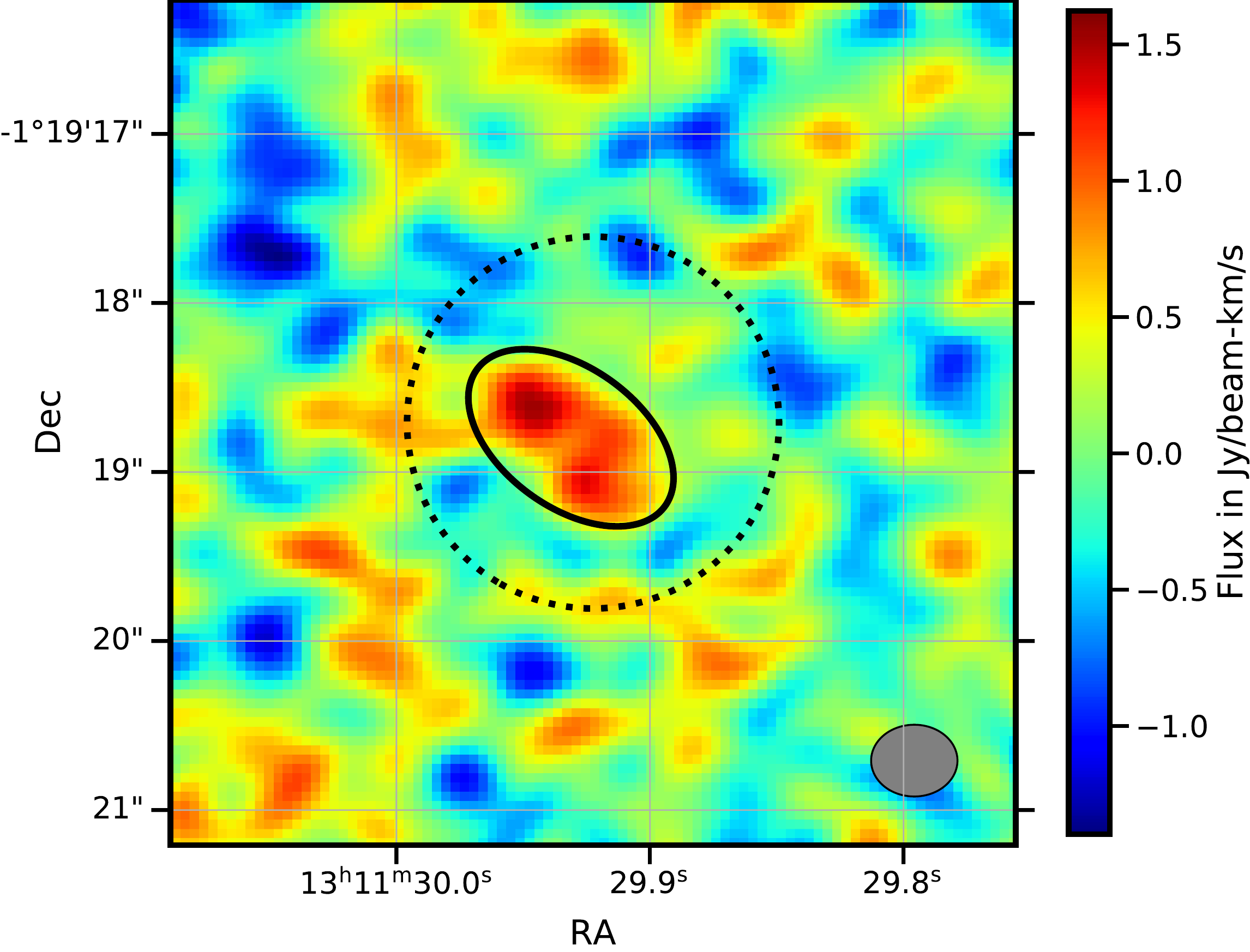}\hfill%
    \includegraphics[height=0.22\textheight, keepaspectratio]{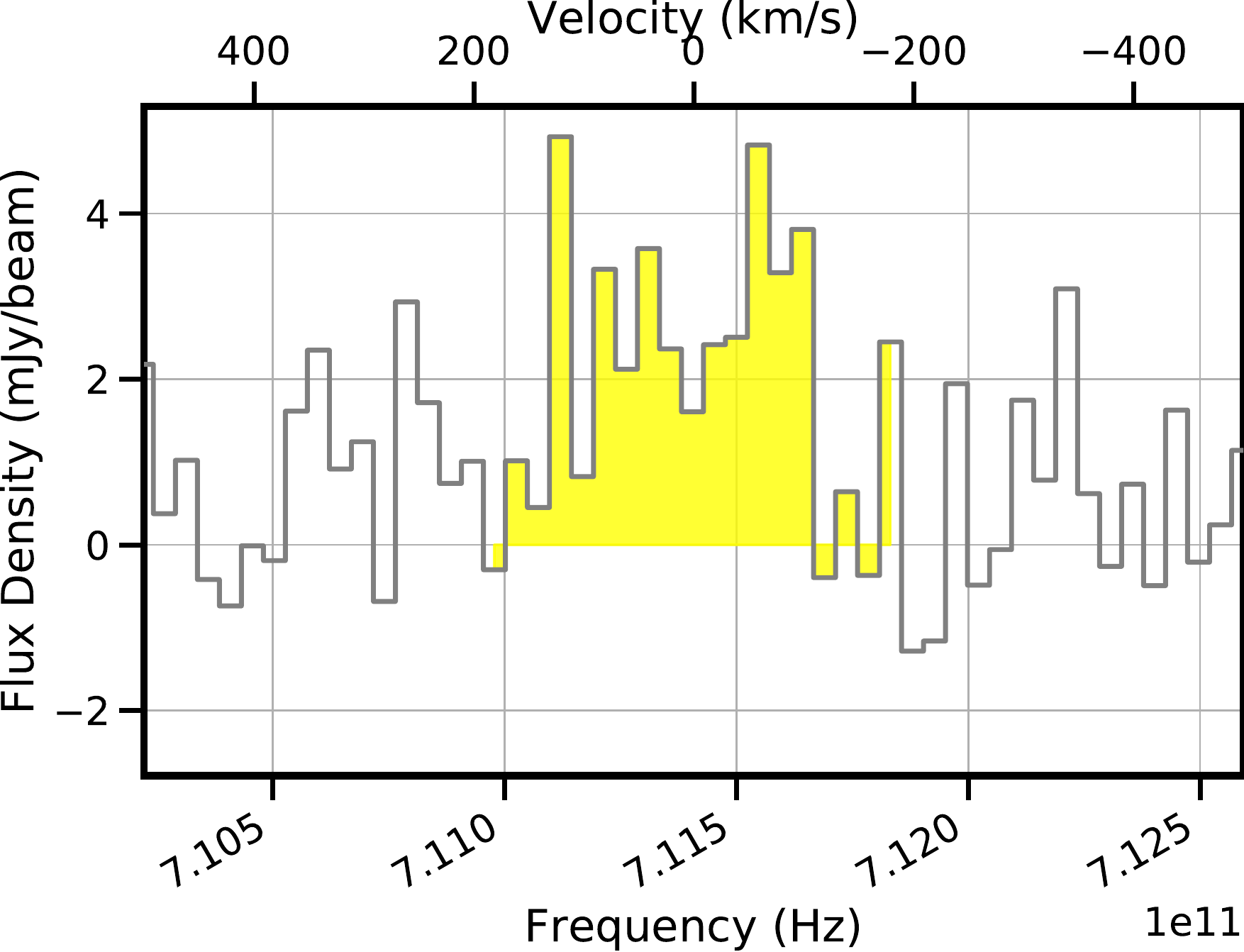}\\
    \includegraphics[width=0.1\textwidth]{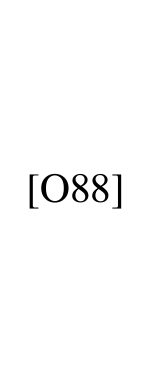}\hfill%
    \includegraphics[height=0.22\textheight, keepaspectratio]{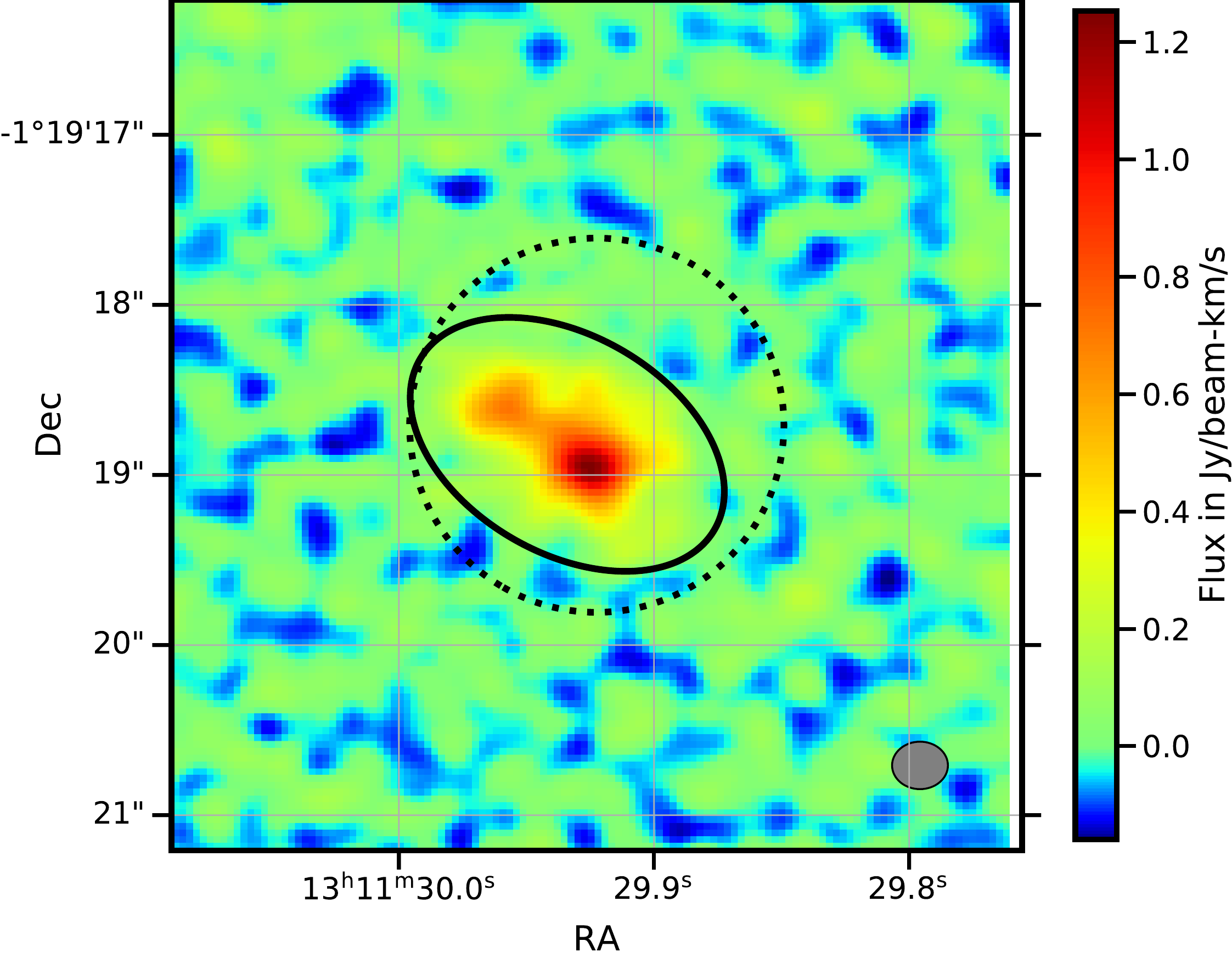}\hfill%
    \includegraphics[height=0.22\textheight, keepaspectratio]{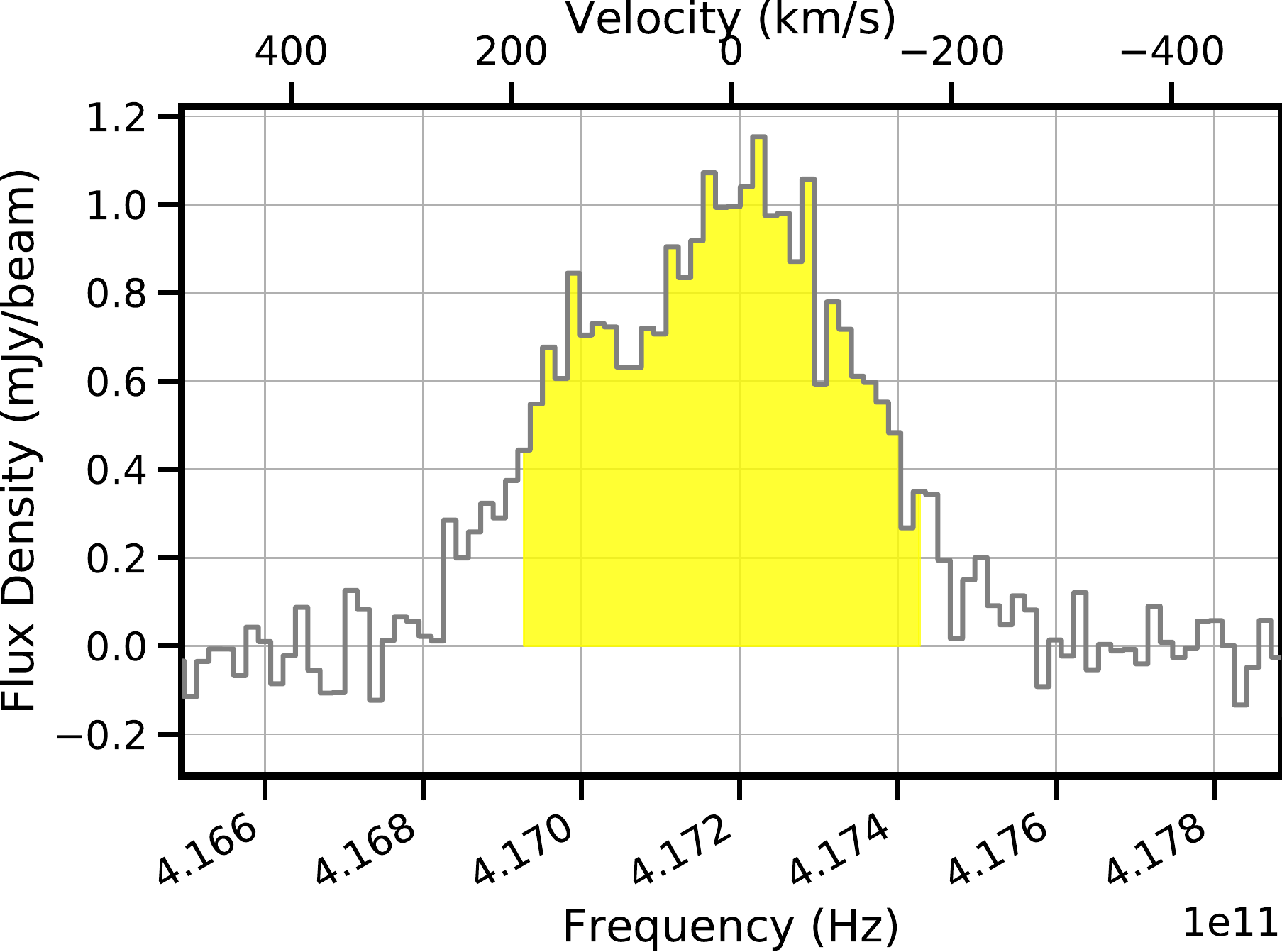}\\
    \includegraphics[width=0.1\textwidth]{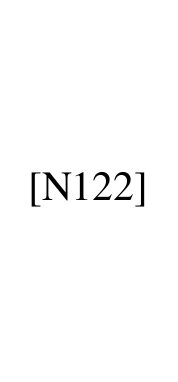}\hfill%
    \includegraphics[height=0.22\textheight, keepaspectratio]{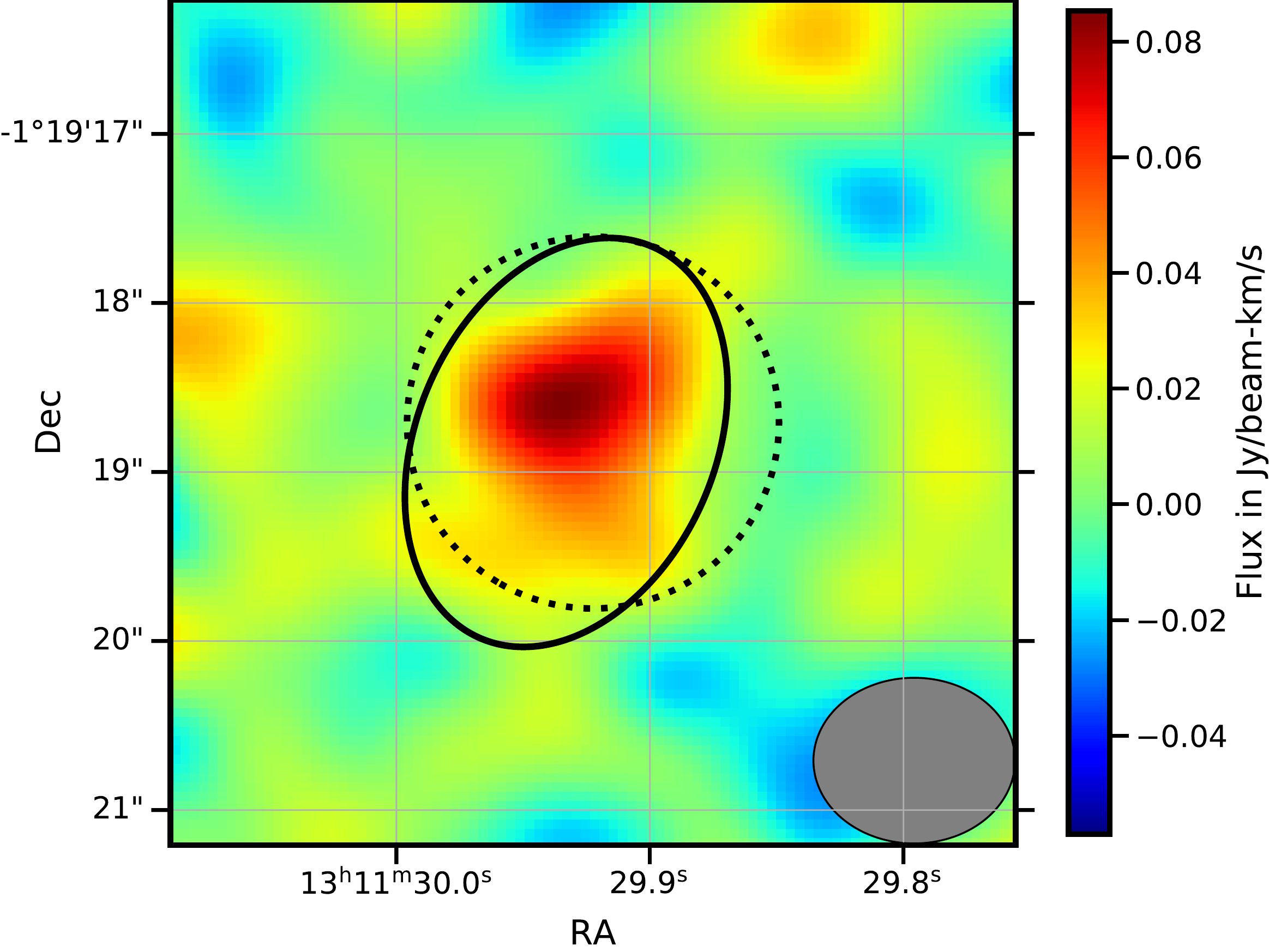}\hfill%
    \includegraphics[height=0.22\textheight, keepaspectratio]{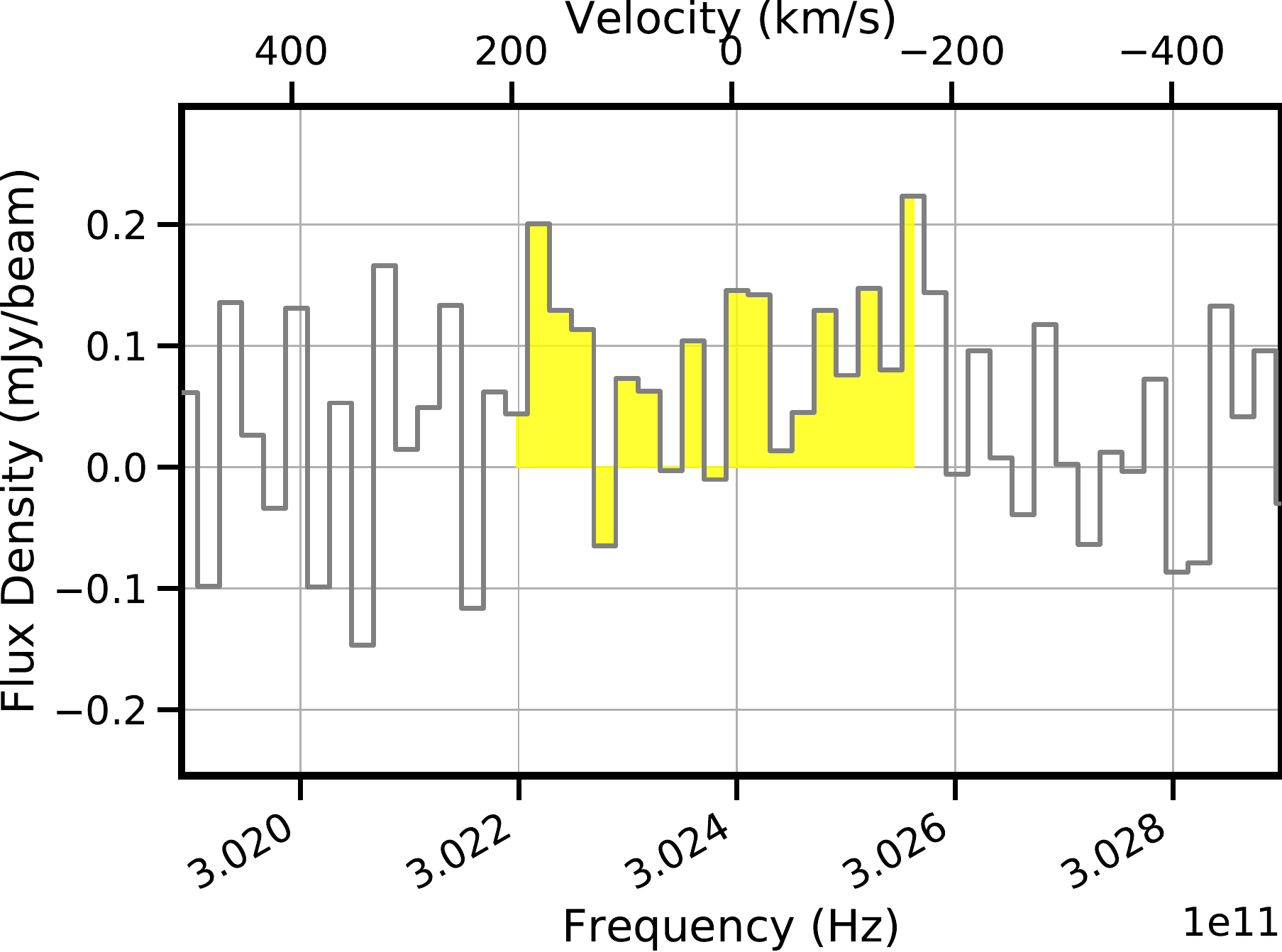}\\
    \includegraphics[width=0.1\textwidth]{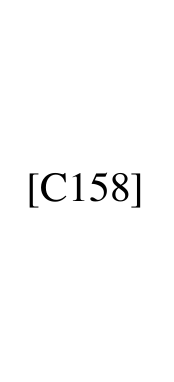}\hfill%
    \includegraphics[height=0.22\textheight, keepaspectratio]{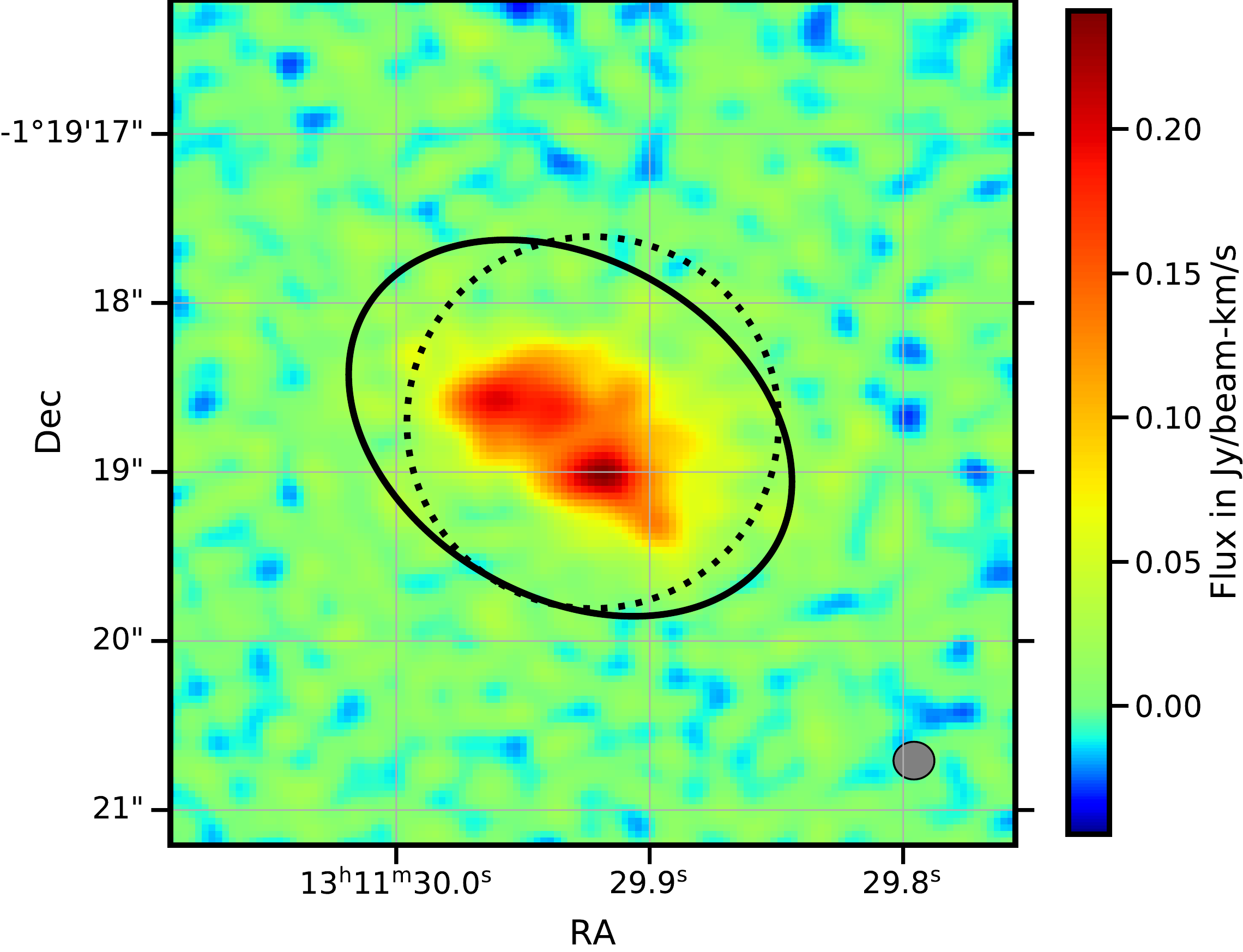}\hfill%
    \includegraphics[height=0.22\textheight, keepaspectratio]{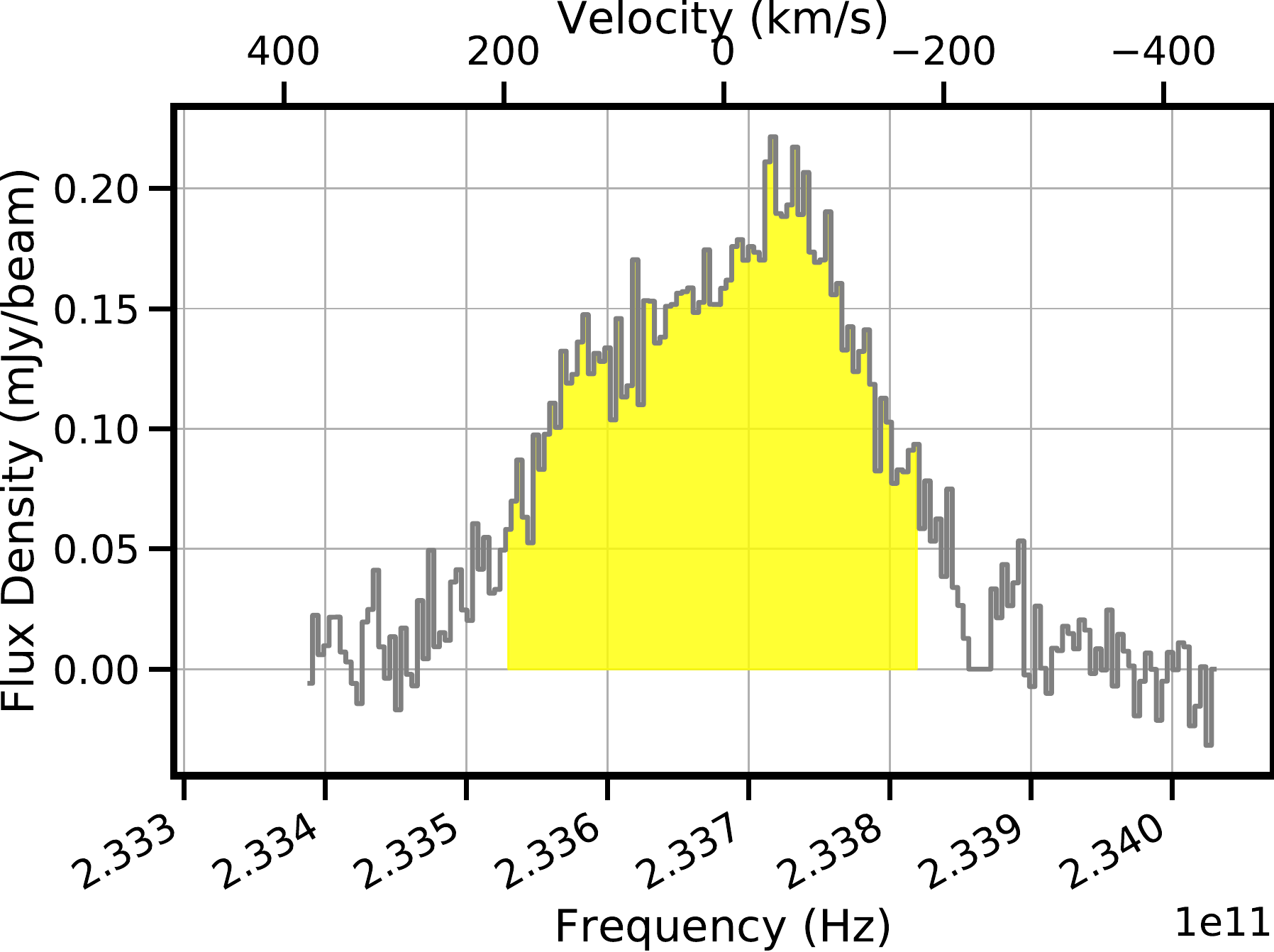}
    \caption{Velocity-integrated moment-0 map (left) and spectrum (right) for all four lines. We show two apertures for each line: the aperture used to extract the spectrum shown as an empty black ellipse, and the aperture used to extract the flux shown as an empty dotted ellipse. The former is customised to each line to extract the best possible spectrum. The latter, common aperture (see Sec.~\ref{sec:flux_measurement}), has the same size and location for all lines to ensure that we use the same physical region to calculate line ratios and estimate metallicity. The beam size is shown by a filled grey ellipse. The highlighted spectral line bins are based on the [C158] line width of -180 to +200 km\,s$^{-1}$. No smoothing is applied.}
    \label{fig:all_mom0_spec}
\end{figure*}

\begin{figure*}
\captionsetup[subfigure]{labelformat=empty, position=top}
\centering
     \begin{subfigure}[b]{0.45\textwidth}
         \centering
         \caption{\large 88\,$\upmu$m continuum}
         \includegraphics[width=\textwidth]{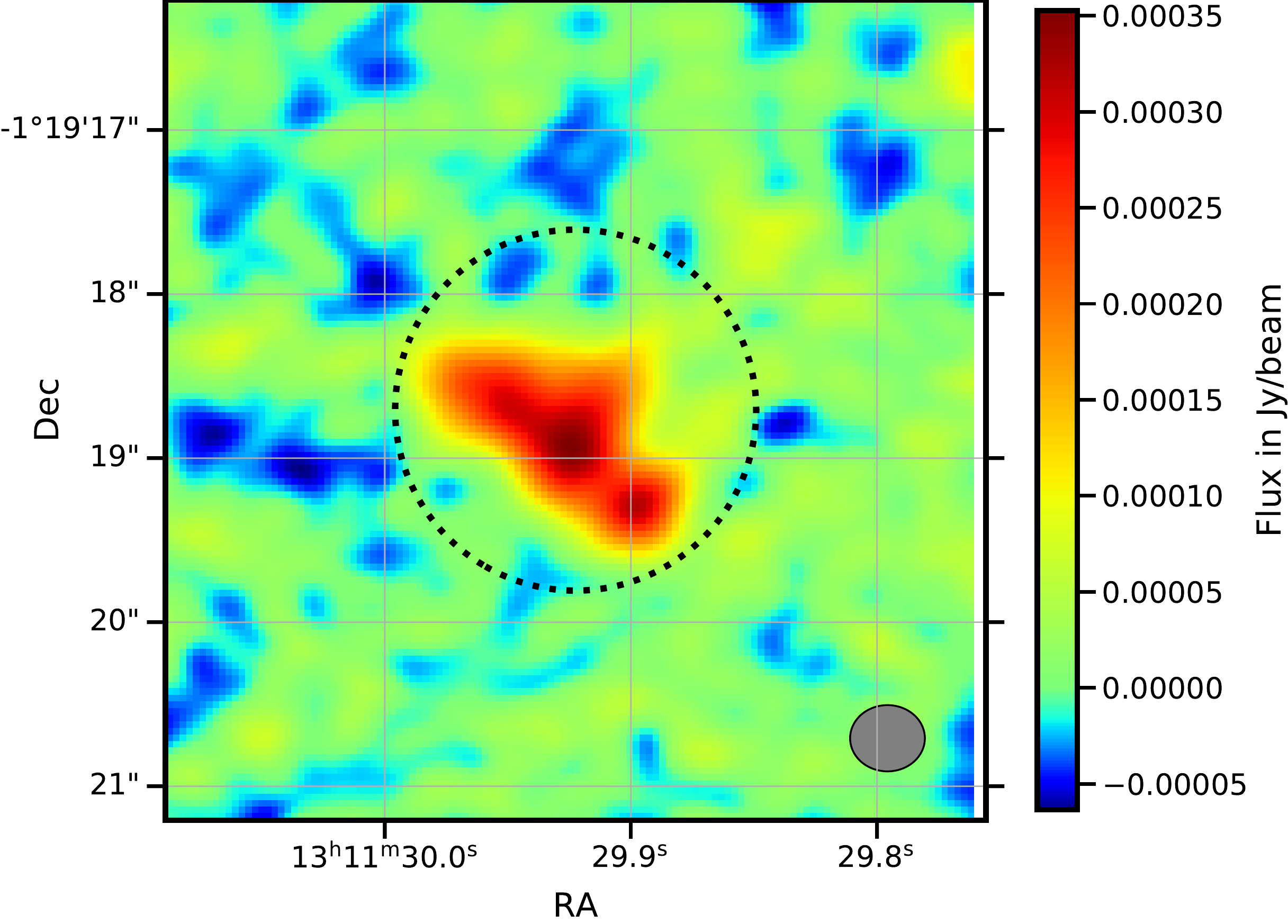}
     \end{subfigure}
     \hfill
     \begin{subfigure}[b]{0.45\textwidth}
         \centering
         \caption{\large 122\,$\upmu$m continuum}
         \includegraphics[width=\textwidth]{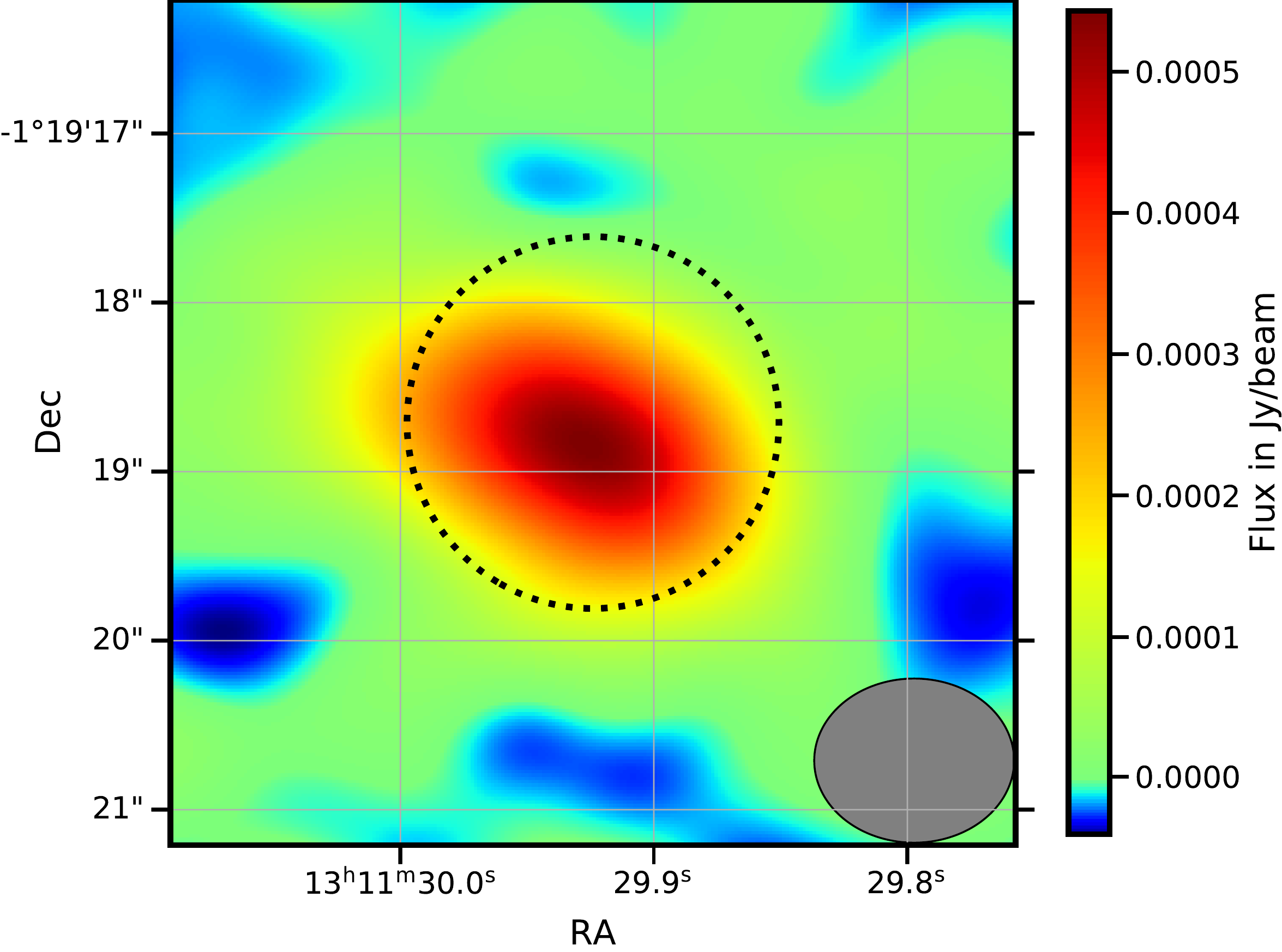}
     \end{subfigure}
    \caption{Continuum maps at 88 and 122\,$\upmu$m. As in Fig.~\ref{fig:all_mom0_spec}, the empty dotted ellipse and the filled grey ellipse show the common aperture used to extract the flux and the beam size respectively.}
    \label{fig:all_cont}
\end{figure*}

\begingroup
\setlength{\tabcolsep}{6pt}
\begin{table}
	\caption{Galaxy-integrated line and continuum measurements for A1689-zD1 using an aperture as described in Sec.~\ref{sec:flux_measurement}. The luminosity has been corrected for lensing, but the flux is uncorrected.}
	\label{tab:zD1}
    \begin{tabular}{@{}cccc@{}} 
    \hline\hline
		Line & $\lambda_{\text{rest}}$ & Flux & Luminosity ($L_\odot$)\\
    \hline
		\oiii & 52\,$\upmu$m & $2.3\pm 1.6$ (Jy\,km\,s$^{-1}$) & $9.6\pm 6.7\times10^{8}$\\
		\oiii & 88\,$\upmu$m & $5.75\pm 0.38$ (Jy\,km\,s$^{-1}$) & $1.40\pm 0.09\times10^{9}$\\
		\nii & 122\,$\upmu$m & $0.09\pm 0.03$ (Jy\,km\,s$^{-1}$) & $1.65\pm 0.51\times10^{7}$\\
		\cii & 158\,$\upmu$m & $3.56\pm 0.07$ (Jy\,km\,s$^{-1}$) & $4.84\pm 0.10\times10^{8}$\\[3pt]
		continuum & 88\,$\upmu$m & $1.72\pm0.13$ (mJy)\\
		continuum & 122\,$\upmu$m & $0.82\pm 0.03$ (mJy)\\
    \hline
	\end{tabular}
\end{table}



In Figure \ref{fig:all_mom0_spec}, 
we show the velocity-integrated moment~0 maps\footnote{produced using the \texttt{spectral-cube} package in Python \citep{ginsburg_radio-astro-toolsspectral-cube_2019}} and spectra for the [N122] and [O52] lines (along with the [C158] and [O88] detections from previous studies). 
The [N122] and [O52] lines show significance levels of \(5.0\sigma\) and \(3.7\sigma\) at the peak pixel, respectively. 
The respective significance is \(3.4\sigma\) and \(3.8\sigma\) in the aperture optimised to each line (Fig.~\ref{fig:all_mom0_spec}), and \(3\sigma\) and \(1.4\sigma\) in the common aperture (see Sec.~\ref{sec:flux_measurement}).
The morphology of the [O52] line is spatially extended (well beyond the beamsize), consistent with the spatial position and rough extent of the rest-frame UV continuum observed with the {\it Hubble Space Telescope} (\emph{HST}) \citep{watson_dusty_2015}.
Given the consistency with \emph{HST}, we conclude that we achieve the first detection of the faint FIR lines of [O52] as well as [N122] at \(z>7\).

\subsection{Flux measurement}
\label{sec:flux_measurement}
To perform a fair photometric comparison by analysing the same regions of the galaxy, we use a common aperture to extract the enclosed flux for the four lines and the underlying continua for [N122] and [O88]. The common aperture was selected to get the best estimate of the weakest lines, [O52] and [N122]. We use this common aperture for all our calculations. To find the best common aperture, we plotted the signal-to-noise ratio (SNR) as a function of increasing aperture radius for both [O52] and [N122]. For [O52], the highest SNR was at $0.''5$ radius beyond which noise began to dominate. For [N122], the optimal aperture radius was around $1.''0$. As the beamsize of the [N122] line was $1.''19\times0.''98$, we chose not to use the optimal [O52] aperture to avoid flux loss in an aperture with diameter smaller than the largest beamsize. Hence, we used a circular aperture with $1.''1$ radius to include most of the [N122] and [O52] flux. We also adopt a common velocity integration range of \([-180:+200]\) km\,s$^{-1}$ to estimate the line flux. This range is based on the \(\sim 2\sigma\) velocity width for the [C158] line as can be seen from the last panel of Fig.~\ref{fig:all_mom0_spec}. 

We use a circular aperture of $1.''1$ radius, centred at RA~=~13:11:29.924 and Dec.~=~$-$01:19:18.710 (J2000). The aperture was chosen to include both the [O52] and [N122] lines, which is slightly larger than the detectable [O52] emission region (see Figs.~\ref{fig:all_mom0_spec} and \ref{fig:all_cont}). This aperture also encompasses the central [C158] and [O88] emission regions. We ensured that the aperture size is not smaller than the beam-size of our worst resolution image ([N122]). The fluxes and corresponding luminosities are shown in Table~\ref{tab:zD1}.

To test whether the difference in resolution affects our flux measurement, we tapered the higher resolution [O88] map to match the lower resolution [O52] and [N122] maps. The fluxes measured were consistent to the values reported in Table~\ref{tab:zD1} within \(\sim 1\sigma\) uncertainty. Additionally, we tested several elliptical and circular apertures that also encompassed all four line emissions, and the results were consistent.

The detection significance of the [O52] and [N122] lines is \(\sim 3.5 \sigma\). To test the significance of the line detection further, we employed a moving spectral window and produced several moment-0 maps with midpoints across the velocity axis. Then we performed a systematic search for off-centre sources in each moment-0 map using a $1.''1$ circular aperture. While for [N122], we found no other sources with \(\geq 3.5 \sigma\) significance, we did find a few of them for [O52]. However, these did not have extended spatial morphologies like the central source. Moreover, [O52] was only used to derive an upper limit (see Sec.~\ref{sec:oiii_ne}), so if the detection significance is lower, our upper limit still holds.

\subsection{Metallicity Constraint}
In this section, we obtain a constraint for the metallicity, $Z$, of A1689-zD1 in a series of steps. We first derive the electron density, $n_e$, using the \oiii line ratio. We then combine this with the 88 to 122\,$\upmu$m dust continuum ratio to derive the ionisation parameter, $U$. Finally, using $U$ and the [O88] to [N122] line luminosity ratio, we constrain $Z$.

The flux ratios could in principle be affected by differential magnification, which in turn depends on the lensing model assumed. However, in this case, since we are calculating integrated galaxy properties in a common aperture and all the lines and continuua are mostly co-spatial, differential magnification is likely to be small, only of the order of a few per cent.

\subsubsection{\texorpdfstring{\oiii}{OIII} ratio}
\label{sec:oiii_ne}
The ratio of the [O52] to [O88] luminosity is independent of both $Z$ and $U$ as both lines originate from the same ion, and of the temperature, because the energy difference between these two states is small compared to the typical gas temperature in the ionising regions of the galaxy. It is therefore a robust probe of $n_e$, up to $10^4$ or even $10^5$\,cm$^{-3}$ \citep{palay_improved_2012, pereira-santaella_far-infrared_2017, zhang_far-infrared_2018,yang_analytic_2020}.

Fig.~\ref{fig:oiii_ne} shows the theoretical relationship between the \oiii line ratio and $n_e$. The [O52] to [O88] ratio for A1689-zD1 in the common aperture is plotted as a horizontal purple line with \(1\sigma\) uncertainty plotted as the corresponding shaded purple region. We derive a nominal value of $n_e \sim$ 55\,cm$^{-3}$ for the electron density. Including the \(1\sigma\) uncertainty on the ratio, we obtain 1 and \(2\sigma\) upper limit of $n_e \lesssim$ 260 and 485\,cm$^{-3}$.

Our density derivation assumes that the gas is optically thin and in thermodynamic equilibrium at a temperature of 10\,000\,K. The upper limit is less than $10^3$\,cm$^{-3}$ for any temperature between 5\,000 and 20\,000K. In the following analysis, we adopt the \(1\sigma\) bound of $n_e \sim$ 260\,cm$^{-3}$ to propagate into our uncertainty calculation.

\begin{figure}
	\includegraphics[width=\columnwidth,clip=]{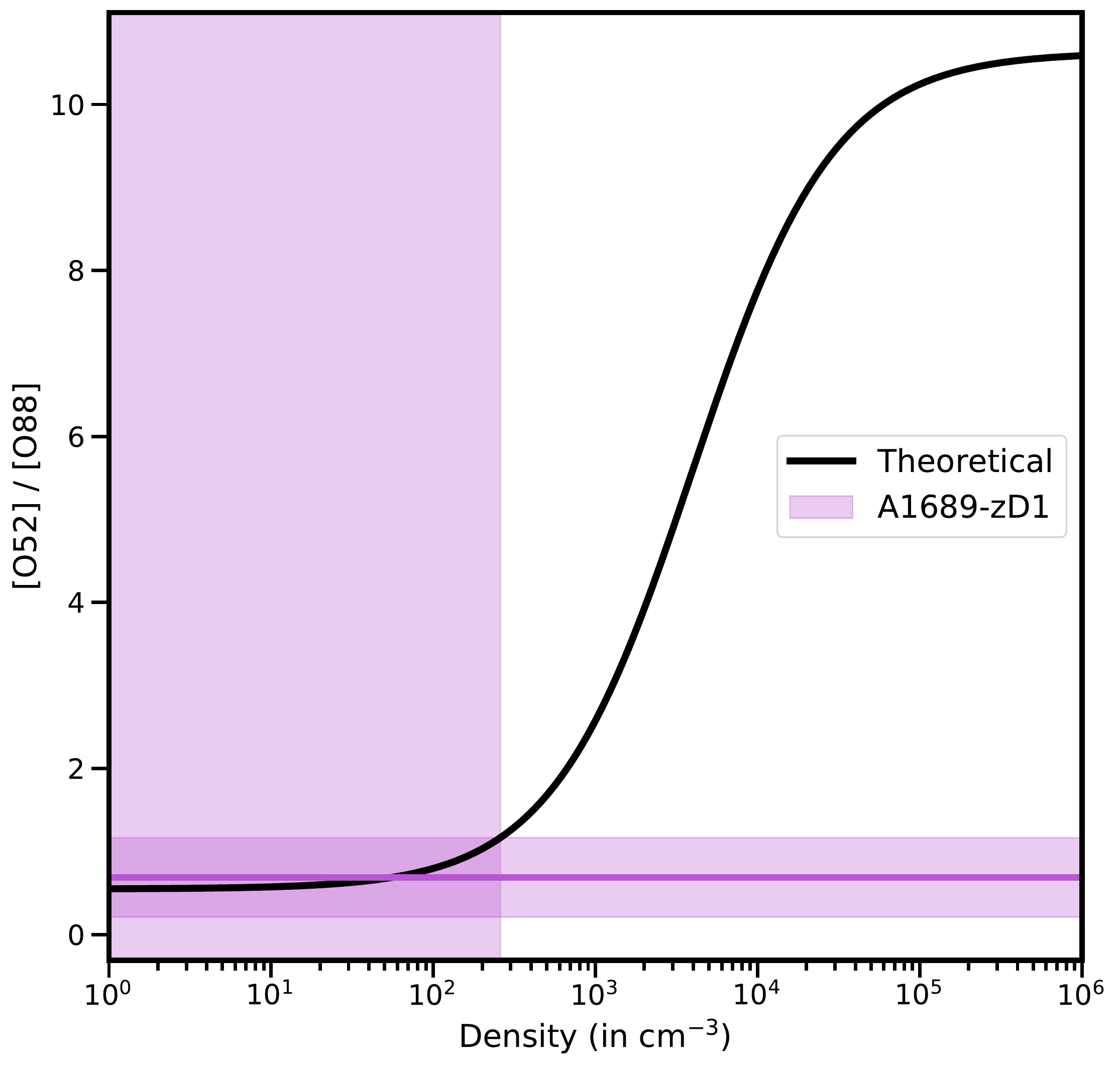}
    \caption{Theoretical relationship between [O52]/[O88] ratio and density shown as a black curve with the ratio for A1689-zD1 along with \(1\sigma\) uncertainty shown in purple. The intersection of the horizontal purple line and black curve gives the nominal density measurement of $n_e \sim$ 55\,cm$^{-3}$, and the intersection of the extreme ends of the horizontal shaded purple region with the black curve gives the uncertainty range on the density. In case of A1689-zD1, we are only able to derive an upper limit of $n_e \sim$ 260\,cm$^{-3}$.}
    \label{fig:oiii_ne}
\end{figure}




\subsubsection{Dust continuum ratio and $U$}
\label{sec:logU}
The ratio of the continuum at 88\,$\upmu$m and 122\,$\upmu$m can be used to constrain $U$, with some dependence on the density \citep{rigopoulou_far-infrared_2018}. We assume a $1\sigma$ density range with an upper bound of 260\, cm$^{-3}$ from the \oiii line ratio and a lower bound of about 10\, cm$^{-3}$ (corresponding approximately to a uniform distribution of \(2\times10^{10}\,M_\odot\) in gas over the galaxy area). While this lower bound is somewhat arbitrary, it is the upper density bound that influences how low the metallicity can be. A lower density would result in higher metallicity and ionisation parameter. In Fig.~\ref{fig:rigo_cont}, we plot the continuum ratio as a function of $\log{U}$ based on CLOUDY modelling over these density bounds from \citet{pereira-santaella_far-infrared_2017}. We show the ratio for A1689-zD1 with \(1\sigma\) uncertainty regions. The extreme values of this uncertainty region are then propagated through the model at the extreme values of the density range derived in Sec.~\ref{sec:o88_nii}. From this, we infer a value of $-1.7 \lesssim \log{U} \lesssim -0.8$ within the \(1\sigma\) uncertainty range.

\begin{figure}
     \includegraphics[width=\columnwidth]{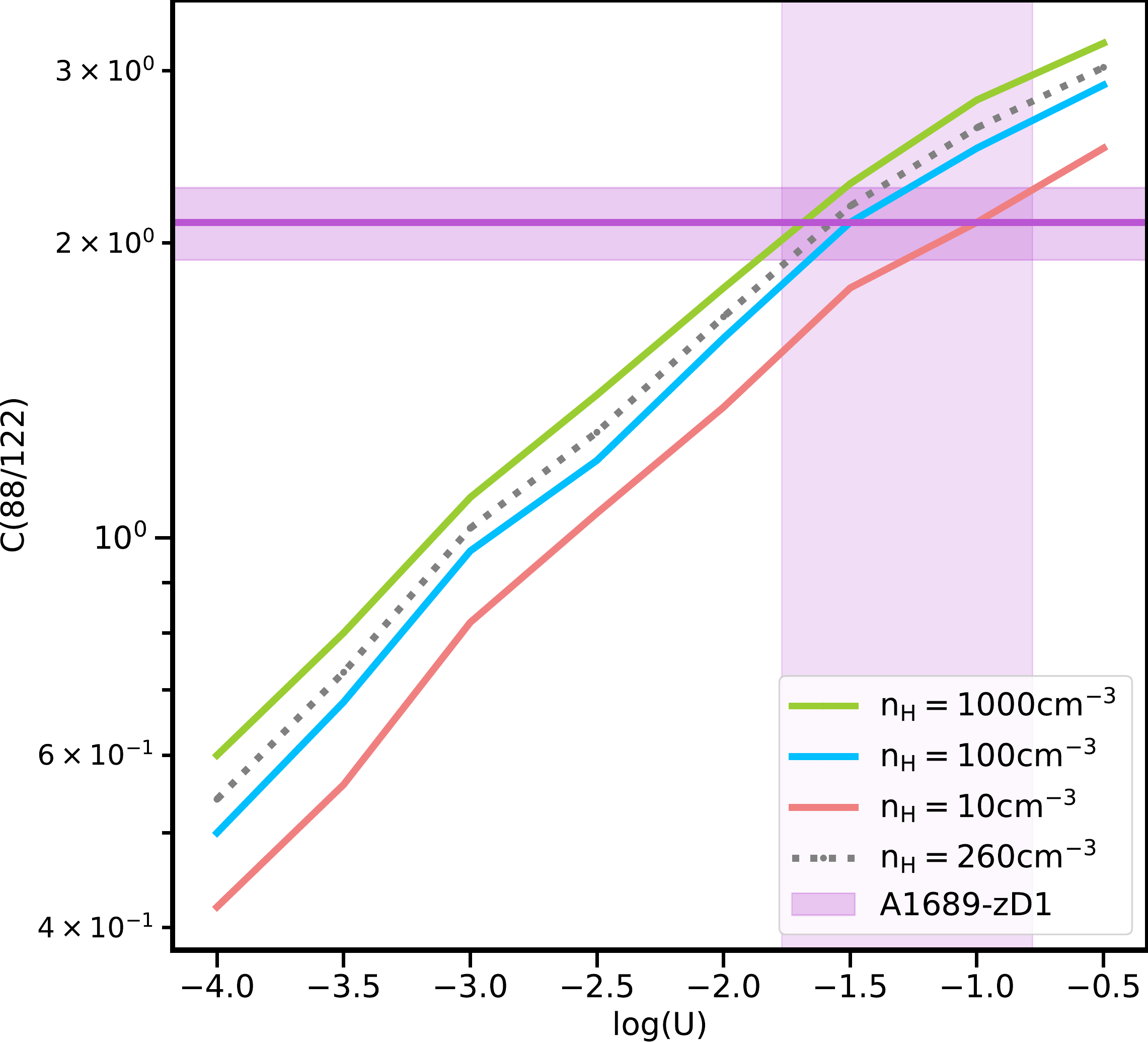}
     \caption[Estimating Z]{[O88] to [N122] continuum ratio as a function of U and $n_H$. Figures from \citealt{rigopoulou_far-infrared_2018} re-created with permission. Measurement for A1689-zD1 is shown as a purple line with the \(1\sigma\) uncertainties depicted as a shaded purple region. As in Fig.~\ref{fig:oiii_ne}, the uncertainty range on $\log{U}$ is given by the intersection of the extreme ends of the horizontal shaded purple region with the model curves corresponding to the extreme values (dotted grey curve for upper limit and red curve for lower limit) of our density estimate from Sec.~\ref{sec:oiii_ne}.}
     \label{fig:rigo_cont}
\end{figure}

\subsubsection{[O88]/[N122] ratio and the metallicity}
\label{sec:o88_nii}

Since the [O88] and [N122] lines have similar critical densities, their ratio is nearly independent of the density. However, it does depend on $Z$ and $U$. Fig.~\ref{fig:rigo_line} plots the ratio as a function of $Z$ for different model tracks of $\log{U}$, once again using the \citet{pereira-santaella_far-infrared_2017} model. While the model does hold beyond $\log{U} > -2$, this parameter space was only explored in their work for galaxies with an active galactic nucleus (AGN). Non-AGN galaxies generally do not have $\log{U} > -2$, but A1689-zD1 appears to be an exception with a high $\log{U}$ despite not having any appreciable AGN activity. We therefore extrapolate the non-AGN \citet{pereira-santaella_far-infrared_2017} model plot to higher values of $U$ to accommodate the measurements for A1689-zD1. Since these extrapolated model values are in agreement with the numbers in the \citet{harikane_large_2020} models presented below, which is also CLOUDY based, and does extend all the way up to $\log{U} = -0.5$, we are confident that the extrapolation is valid.

As before, the ratio for A1689-zD1 is indicated with the \(1\sigma\) uncertainty regions. Once again, we derive the uncertainty range on metallicity by propagating the extreme values of the uncertainty region on the [O88]/[N122] ratio through the model curves at the extreme values of the $\log{U}$ measurements from Sec.~\ref{sec:logU}. We thus find $0.9 \lesssim Z/Z_\odot \lesssim 1.3$. As mentioned in Sec~\ref{sec:logU}, if we were to allow lower densities, $U$ and in turn, $Z$, would be higher.

\begin{figure}
     \includegraphics[width=\columnwidth]{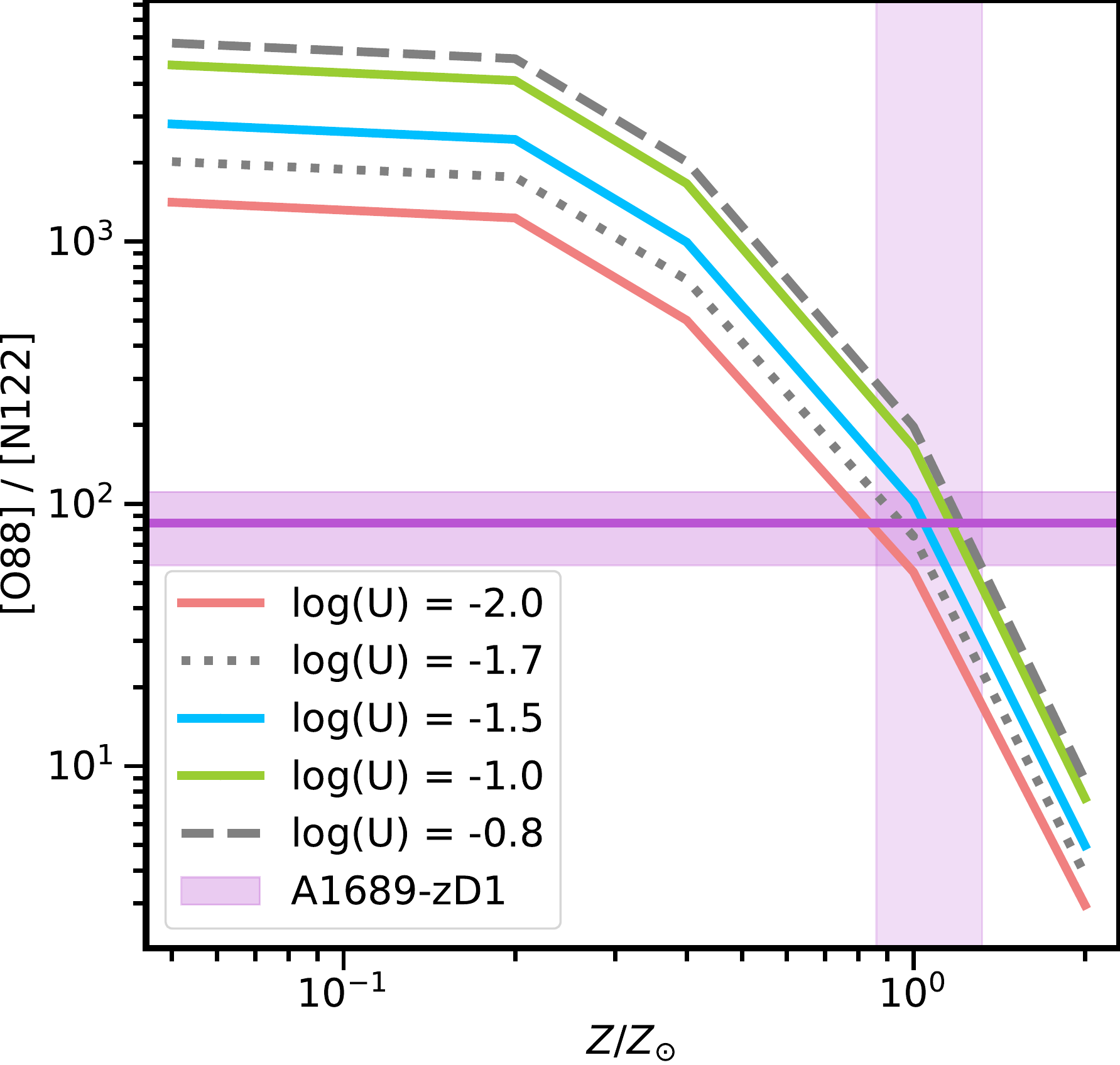}
     \caption[Estimating U]{[O88] to [N122] line ratio as a function of U and Z (Adapted from \citealt{rigopoulou_far-infrared_2018} and extrapolated above -2.0). Measurements for A1689-zD1 are shown in purple, just as in Fig.~\ref{fig:rigo_cont}. Also as in Fig.~\ref{fig:rigo_cont}, the uncertainty range on $\log{U}$ is given by the intersection of the extreme ends of the horizontal shaded purple region with the model curves corresponding to the extreme values (dotted grey curve for lower limit and dashed grey curve for upper limit) of our $\log{U}$ estimate from Sec.~\ref{sec:logU}.}
     \label{fig:rigo_line}
     \end{figure}

For comparison, we use models from \citet{harikane_large_2020} with metallicity-dependent N/O and C/O ratios. These are plotted in Fig.~\ref{fig:harix}. The model assumes that the nitrogen-to-oxygen abundance ratio depends on the metallicity due to secondary nucleosynthesis (see Sec.~\ref{sec:N_excess}). This in turn uses the relation presented in \citet{kewley_using_2002} in the same manner as \citet{nagao_metallicity_2011}. With this model, we find $Z/Z_\odot \sim 1$ to $2$, and $\log{U} \sim -0.5$ to $-2$ (with $n_{\rm H} \sim 10$ to $100\,$cm$^{-3}$) roughly consistent with the estimates based on \citet{rigopoulou_far-infrared_2018}.

\begin{figure}
     \includegraphics[width=\columnwidth]{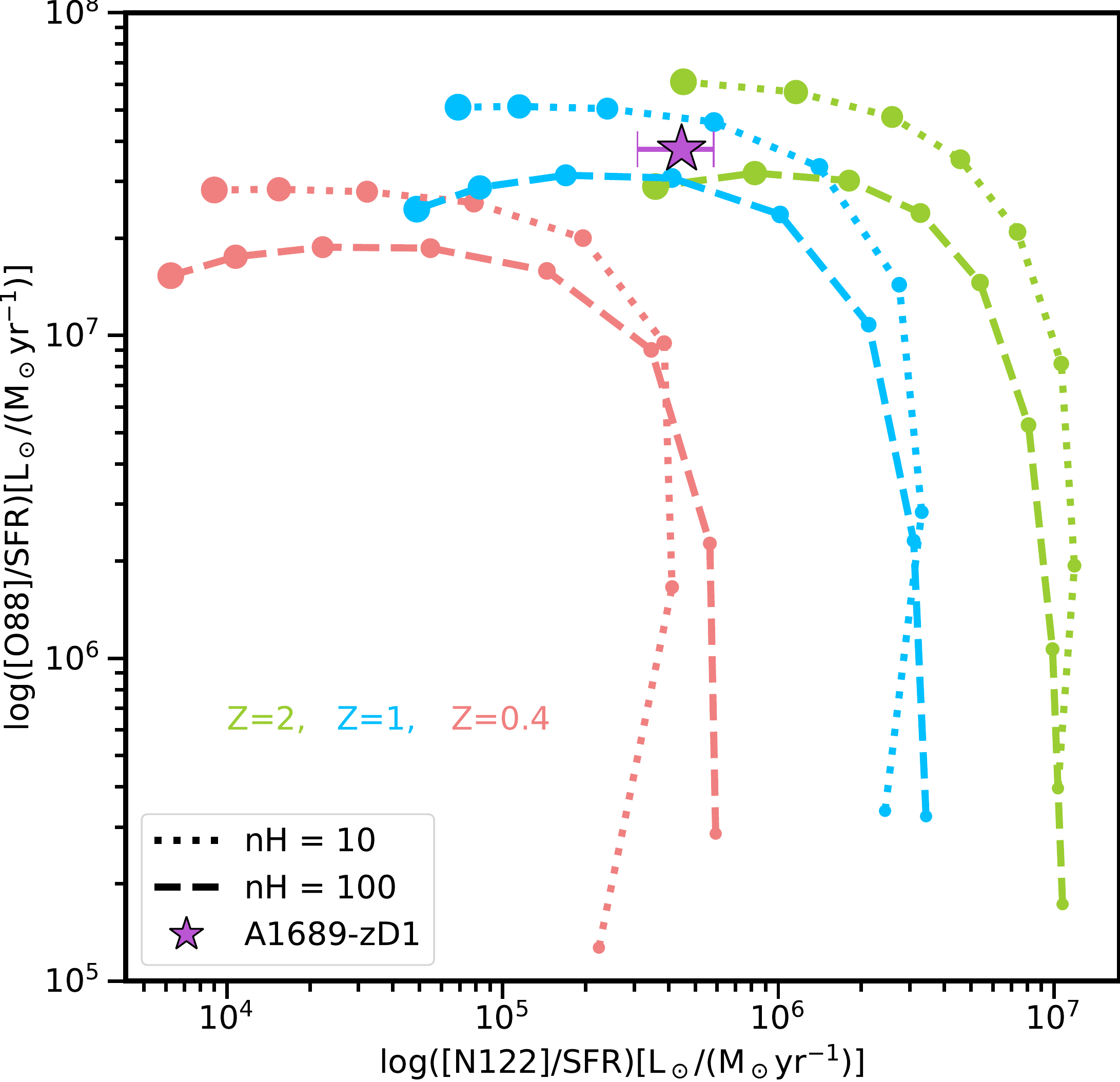}
     \caption[Harikane Z estimate]{Model curves at various metallicities and densities for the [O88] luminosity/SFR ratio as a function of the [N122] luminosity/SFR ratio, based on \citealt{harikane_large_2020}. Increasing marker size represents an increase in $\log{U}$, within the range -4 to -0.5. The measurements for A1689-zD1 are shown as a purple star with uncertainties.}
     \label{fig:harix}
\end{figure}

Despite both models being CLOUDY-based, the slight difference in metallicity estimate may arise from the different assumptions made in each one. For instance, \citet{harikane_large_2020} assumes a Chabrier initial mass function (IMF) and \citet{pereira-santaella_far-infrared_2017} assumes a Kroupa IMF. In addition, we use a modified \citet{harikane_large_2020} model with metallicity-dependent nitrogen abundance, but a similar modification was not made for \citet{rigopoulou_far-infrared_2018} model. Regardless of these differences, metallicities significantly below the solar value do not reproduce our line ratios with either the \citet{rigopoulou_far-infrared_2018} or \citet{harikane_large_2020} model.

Nonetheless, there is some uncertainty associated with the model curves presented here. Firstly, there is some scatter in the N/O abundance to metallicity conversion \citep[e.g.][]{liang_oxygen_2006}. Additionally, CLOUDY model curves have a model uncertainty of the order 10--20 per cent as discussed in \citet{pereira-santaella_far-infrared_2017}, comparable to metallicity models based on optical emission lines. Although, as discussed in e.g. \citet{croxall_toward_2013}, models relying on FIR lines remove the heavy dependence on temperature which plagues optical emission lines.

\subsection{[O88]/[C158] ratio and the PDR covering fraction}
\label{sec:pdr}
The ionisation energies of \oiii (35.1\,eV) and [N122] (14.5\,eV) are higher than that of H (13.6\,eV), whereas the ionisation energy of [C158] (11.2\,eV) is lower than that of H. Hence, the [C158] emission comes from the cold atomic components, photo-dissociation regions (PDR), and \hii regions, whereas the emission from the other three lines comes exclusively from the \hii regions. Therefore, the ratio of [C158] to any of the other three lines can be used to estimate the PDR covering fraction, i.e.\ the extent of the ionised H gas compared to the neutral H gas \citep[e.g.][]{cormier_herschel_2019,harikane_large_2020}.

In Fig.~\ref{fig:haripdr}, we plot model curves for [C158] luminosity assuming PDR covering fractions of 0 and 1. The measurements for A1689-zD1 favour a model with PDR fraction close to 1, i.e.\ dominated by neutral atomic gas.

\begin{figure}
     \includegraphics[width=\columnwidth]{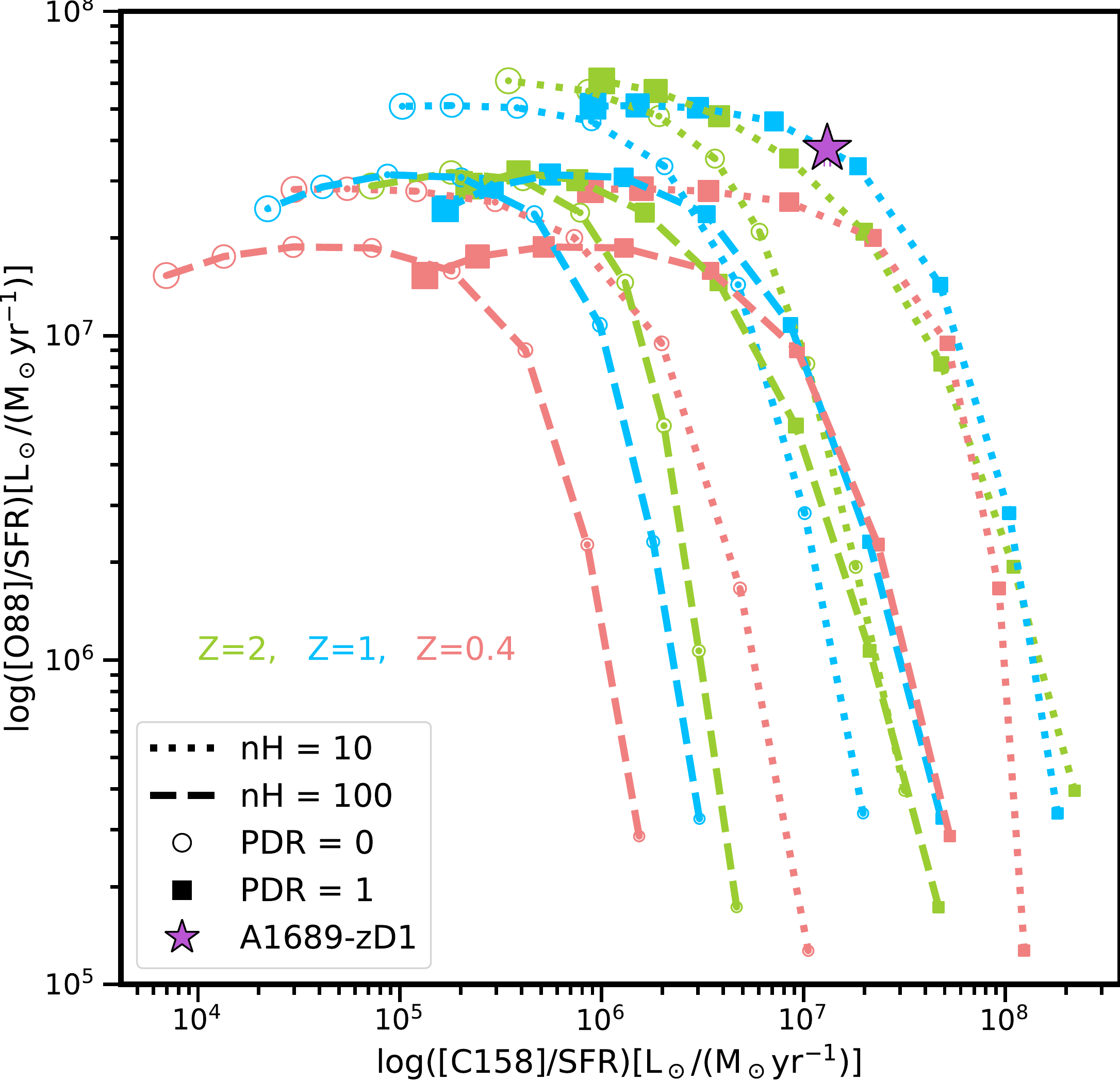}
     \caption[Estimating PDR fraction]{Model curves based on \citealt{harikane_large_2020} for the [O88] luminosity/SFR ratio as a function of the [C158] luminosity/SFR ratio, assuming a PDR covering fraction of 0 (open circles), and 1 (filled squares). The $\log{U}$ range is the same as in Fig.~\ref{fig:harix}, and measurements for A1689-zD1 are shown as a purple star just as in Fig.~\ref{fig:harix}.}
     \label{fig:haripdr}
\end{figure}

\subsection{Dust-to-metals ratio}
\label{sec:dtm}
The total gas mass for A1689-zD1 is based on the sum of the atomic and molecular masses. We determine the atomic gas mass from the relation between the [C158] line luminosity, metallicity and the atomic gas mass from \citet{heintz_measuring_2021}. We find M\(_\mathrm{H\,\textsc{i}} = 1.7_{-0.5}^{+0.7}\times10^{10}\)\,M\(_\odot\). This includes the scatter in the relation and the statistical error added in quadrature.

Assuming most of the gas is in the atomic phase, the total gas mass is between 1.2 and \(2.4\times10^{10}\)\,M\(_\odot\). Using a solar metal fraction of about 1/100 and a dust mass of \(1.7^{+1.3}_{-0.7}\times10^{7}\)\,M\(_\odot\) \citep[][]{bakx_accurate_2021}, the corresponding dust-to-metals mass ratio (DTM) for A1689-zD1 is around 0.1. If there is significantly more gas in the system, say in the molecular phase, this will make the DTM even lower.


\section{Discussion}

Several striking features are derived from the line ratios reported here. First, the metallicity is close to solar, which deviates strongly from the expected mass-metallicity evolution towards high-$z$.

Second, the nitrogen abundance is in excess of oxygen, indicating that the star-formation in the galaxy must be old enough to have produced secondary nitrogen through intermediate mass stellar envelopes, i.e.\ at least 250\,Myr \citep{henry_cosmic_2000}, pushing the formation age of the system back to \(z>10\).

Third, though this system is known as a prototypical dusty normal galaxy at this epoch, it seems to be deficient in dust compared to its total metal content.

\subsection{Evolution of metal abundance}
The most reliable metallicity estimates for star-forming galaxies come from back-lighting absorption studies, e.g.\ gamma-ray burst (GRB) afterglows. All show metallicities substantially below 0.1 solar
\citep{salvaterra_high_2015} at \(z\sim6\). With some assumptions on density and temperature, \citealt{jones_mass-metallicity_2020} have estimated metallicities for a handful of \(z>7\) galaxies using the relative strength of the [O88] line to the total SFR and find metallicities ranging from 8--36 per cent of the solar value. Using a similar method to the one used here, but assuming $\log{U}$, \citet{novak_alma_2019} find solar or possibly super-solar metallicity for the ISM of the host galaxy of the quasar J1342+0928 at \(z=7.54\), demonstrating that such a high metallicity is not unique at \(z>7\).
Over the next 13\,Gyr, if such a high metallicity galaxy is to increase its mass, it must do so mainly via dry mergers and not through a lot of star formation which would lead to supernova explosions that would substantially increase the metal content.

\subsubsection{The fundamental metallicity relation}
\label{sec:fmr}
Galaxies up to \(z \sim 2.5\) lie on a plane in 3D space spanning $M_\ast$, $Z$, and SFR. While there appears to be no evolution between local Sloan Digital Sky Survey (SDSS) galaxies at \(z \sim 0\) and those at \(z \sim 2.5\), there is some evolution above \(z \sim 2.5\) \citep{mannucci_fundamental_2010}. We do not know how early these relationships are set up in galaxies, but galaxies at \(z \sim 3\) appear to lie 0.6\,dex below the metallicity prediction of the lower-$z$-calibrated fundamental metallicity relation (FMR) from \citet{mannucci_fundamental_2010}. Other studies have also found an evolution of the FMR relation with redshift \citep[e.g.][]{stott_fundamental_2013,torrey_similar_2018,sanders_mosdef_2021}.


Given the stellar mass and SFR for A1689-zD, its metallicity is substantially higher than the \(z \sim 3\) FMR by \(\sim 1.25\)\,dex, and even the \(z \sim 0\) FMR by \(\sim 0.6\)\,dex. Hence, the measured metallicity of A1689-zD1 is inconsistent with the trend suggested by \citet{mannucci_fundamental_2010} by about an order of magnitude in metallicity. A revised \(z \sim 0\) FMR parameterisation was presented by \citet{curti_mass-metallicity_2020}, but A1689-zD1 deviates from this relation as well by about 1\,dex.

The reason for the deviation may be an inaccurate estimate of the stellar mass. The current value of \(2\times10^9\,\mathrm{M}_\odot\) is determined from rest-frame optical SED fitting, which could be heavily dust-obscured. 
The SFR of A1689-zD1 is more than 90 per cent obscured. While the obscuration of the stellar mass is unlikely to be as high as this, it could still be substantial. With a stellar mass of \(10^{10}\,\mathrm{M}_\odot\), i.e.\ a factor of 80 per cent obscuration of the stellar mass, the deviation from the \(z\sim0\) FMR decreases to only 0.1\,dex (although the deviation from the \(z\sim3\) FMR is still 0.7\,dex). A stellar mass at least as large as the gas mass is required to produce all the metals in a solar metallicity system assuming a Chabrier or Kroupa IMF. Therefore such a high stellar mass is reasonable. However, it is hard to imagine dust obscuration much greater than this.

Another potential reason for the discrepancy might be the assumed N/O ratio in our models based on \citet{rigopoulou_far-infrared_2018} and \citet{harikane_large_2020}. Both models assume the relation between the N/O ratio and metallicity calibrated in the local universe. However, we do not know the N/O-metallicity relation at \(z\sim7\). Some studies \citep[e.g.][]{queyrel_integral_2009, yabe_subaru_2015} report a possible increase of the N/O ratio at fixed metallicity at redshift \(z\sim1.5\), while others do not at \(z\sim2\) \citep{kojima_evolution_2017}. If the N/O ratio does evolve, the estimated metallicity would decrease, and become more consistent with the fundamental metallicity relation.

\subsection{Nitrogen excess and the age of A1689-zD1}
\label{sec:N_excess}
The metallicity estimate derived from the [O88]/[N122] ratio depends on the overabundance of N with respect to O (\citealt{rigopoulou_far-infrared_2018,pereira-santaella_far-infrared_2017}), a consequence of secondary nitrogen production that only becomes dominant at $ Z/Z_\odot\gtrsim0.25$ \citep[e.g.][]{vincenzo_nitrogen_2016,pilyugin_abundance_2014}.

\citet{henry_cosmic_2000} argue that the secondary production of nitrogen  principally occurs in the asymptotic giant branch (AGB) phase of intermediate mass stars (4--8\,M$_\odot$), while O and C production continues to be dominated by high mass stars or type\,II supernovae. This leads to an increase in the N/O ratio with increasing abundance above $Z/Z_\odot\gtrsim0.25$. However, it also introduces a delay of about 250\,Myr, the main-sequence lifetime of these intermediate mass stars, before the N/O ratio increase can occur.

The fact that we observe a relatively low [O88]/[N122] ratio, and from it infer a metallicity significantly above 0.25 solar, suggests that the stellar age of this galaxy is at least several hundred million years. The galaxy must have therefore started forming stars at \(z>10\). This is somewhat at odds with the stellar age inferred from the SED analysis ($\sim80$\,Myr). However, as argued above, much of the stellar mass may be completely obscured.

The stellar mass inferred from the SED fitting is only detected as far as 4.6\,$\mu$m, corresponding roughly to the $V$-band in the rest-frame. As indicated by the 80--90 per cent obscured SFR fraction, we can infer that the average extinction to most sightlines in the galaxy is high. As also indicated by the relatively modest extinction in the UV-bright parts of the galaxy, the dust distribution is likely very patchy, with some low obscuration regions, and most of the galaxy completely extinguished (for example, a dust mass of $2\times10^7$\,M$_\odot$ spread over about 1\,kpc$^2$ gives an $A_V\sim200$ \citealt{watson_galactic_2011}). This suggests that the stellar mass of $2\times10^9\,$M$_\odot$ inferred from the SED is only a lower bound, and the real stellar mass could be up to an order of magnitude higher. A stellar mass of $10^{10}\,$M$_\odot$ would lead to a characteristic age of about 300\,Myr, consistent with the age required for the N/O overabundance.

We therefore suggest two things from these considerations: that A1689-zD1 may have a much higher stellar mass than previously inferred, and that its stellar age is $\gtrsim300$\,Myr. Hence, it must have started forming stars at a significant rate at \(z>10\). This is consistent with the claims for significant \(z>10\) star-formation inferred from candidate \(z \sim 9\) galaxies \citep[e.g.][]{laporte_probing_2021}.

\subsection{Low dust formation efficiency at z > 7}

With a metallicity estimate in hand, we can address the question of the dust-formation efficiency of a galaxy at \(z\gtrsim7\) for the first time. We calculated in Sec.~\ref{sec:dtm} a dust-to-metals mass ratio of about 0.1. This is significantly lower than the MW value. For a MW gas-to-dust ratio of 150 \citep{galliano_interstellar_2018}, suitable for galaxies close to solar metallicity, the corresponding dust mass in A1689-zD1 for its inferred gas mass should be \(1\text{--}2\times10^{8}\)\,M\(_\odot\), an order of magnitude higher than the measured value of \(1.7^{+1.3}_{-0.7}\times10^7\)\,M\(_\odot\).

Although there have been suggestions that the formation of dust may be less efficient at low metallicity \citep{de_cia_dust-depletion_2016,galliano_interstellar_2018}, at solar metallicity, dust-formation efficiency is observed to be high. In other words, the DTM is expected to be close to the MW value of 0.5 at metallicities at least down to about 0.3\,dex below solar, as is observed in low redshift galaxies \citep{de_vis_systematic_2019}. The small number of estimates from emission-line galaxies at \(z\sim2\) \citep{shapley_first_2020} suggest that the DTM is constant at about 0.5 to that redshift at high metallicity too, with the caveat that for those galaxies the molecular mass was assumed to represent all the gas. Estimates of the DTM for systems detected in absorption indicate both a dependence on metallicity \citep{de_cia_dust-depletion_2016} and no metallicity dependence \citep{zafar_metals--dust_2013,wiseman_evolution_2017}.

Regardless, A1689-zD1 is therefore a unique case, indicating a low dust formation efficiency in spite of the high metallicity. The low dust formation efficiency could point to a gas mass much lower than we have inferred, indicating a breakdown of the [C158]-H\textsc{i} relation of \citet{heintz_measuring_2021}, or a dust formation timescale longer than the stellar mass build-up or nitrogen-enrichment timescale of $\sim250$ million years, or possibly a lack of high-emissivity dust originating in e.g.\ AGB stars. One caveat here is indeed the uncertainty in the emissivity of the dust. Local studies of the dust emissivity or dust mass absorption coefficient suggest lower emissivity by up to a factor of two in high density or higher temperature environments \citep[][]{bianchi_dust_2022, clark_first_2019}. This factor would alleviate some of the tension we observe here, though not eliminate it entirely.

Another way to resolve the tension would be to reduce the inferred metallicity. Reducing the metallicity by a factor of several, coupled with lowering the dust emissivity, could be enough to replicate the MW DTM. However, this would require either that the [N122] line luminosity is over an order of magnitude lower than our estimate or that the N/O line ratio-to-metallicity conversion \citep[][]{rigopoulou_far-infrared_2018,harikane_large_2020} is very different at this redshift (see Sec.~\ref{sec:fmr})



\subsection{[C158] deficit and the initial mass function}
We find no [C158] deficit \citep[e.g.][]{hodge_high-redshift_2020} in A1689-zD1, similar to some other massive galaxies at \( z\sim7\) \citep[e.g.][]{capak_galaxies_2015,schaerer_alpine-alma_2020,schouws_alma_2022}. \citealt{katz_nature_2022} claim that the deficit comes from low C/O abundance at high redshift, which in turn arises from enrichment by low metallicity core-collapse supernovae with a top-heavy IMF with no AGB stars to provide carbon. Since most AGB stars take $\gtrsim1$ billion years to contribute substantially to the ISM, the presence of [C158]-bright sources at \(z\sim7\) militates against the hypothesis of a top-heavy IMF with carbon-deficient supernovae.

\subsection{Metallicity variation across the galaxy}
As the SNR and spatial resolution of the [N122] data is much lower than that of the other lines, we could not create a resolved metallicity map. However, considering 
the fact that the [C158] and dust emission are stronger to the northwest side while the \emph{HST} emission is stronger to the southeast (Knudsen et.~al.(in prep.)), the metallicity may vary across the galaxy. A distinct difference in the metallicity between the major components measured with higher SNR measurements could indicate that the system was in the process of merging \citep{knudsen_merger_2017,wong_alma_2022}.

\section{Conclusions}

We have measured [O52] and [N122] for the first time in a reionisation-epoch galaxy. These measurements, coupled with previous measurements of [O88] and [C158], and several dust continuum detections, have allowed us to determine the electron density and metallicity of the galaxy, subject to modelling uncertainties.

A1689-zD1 appears to have approximately solar gas-phase metallicity, remarkably high and unusual for a normal galaxy at this epoch. The excess of nitrogen to oxygen indicates that the star-formation in this galaxy started at least 250\,Myr earlier, i.e.\ at \(z>10\). The galaxy also appears to be atomic gas dominated, and to have a low dust-to-gas ratio for its metallicity, possibly hinting at a low efficiency for dust production in galaxies at this epoch.



\section*{Acknowledgements}

We thank the anonymous referee for insightful comments, and Shengqi Yang and Luca Di Mascolo for helpful discussions.

This paper makes use of the following ALMA data: ADS/JAO.ALMA\#2013.1.01064.S, ADS/JAO.ALMA\#2015.1.01406.S, ADS/JAO.ALMA\#2016.1.00954.S, ADS/JAO.ALMA\#2017.1.00775.S, ADS/JAO.ALMA\#2019.1.01778.S. ALMA is a partnership of ESO (representing its member states), NSF (USA) and NINS (Japan), together with NRC (Canada), MOST and ASIAA (Taiwan), and KASI (Republic of Korea), in cooperation with the Republic of Chile. The Joint ALMA Observatory is operated by ESO, AUI/NRAO and NAOJ. The National Radio Astronomy Observatory is a facility of the National Science Foundation operated under cooperative agreement by Associated Universities, Inc.

DW and SF are supported in part by Independent
Research Fund Denmark grant DFF-7014-00017. The Cosmic Dawn Center is funded by the Danish National Research Foundation under grant number 140. KK acknowledges support from the Knut and Alice Wallenberg Foundation. FR acknowledges support from the European Union’s Horizon 2020 research and innovation program under the Marie Sklodowska-Curie grant agreement No. 847523 ‘INTERACTIONS’ 

\section*{Data Availability}
The data used in the paper are available in the ALMA archive at https://almascience.nrao.edu. The derived data and models generated in this research will be shared on reasonable request to the corresponding author.

\bibliographystyle{mnras}
\bibliography{references}

\bsp	
\label{lastpage}
\end{document}